\documentclass[letterpaper, 10 pt, conference]{ieeeconf}  

\usepackage{graphicx}
\usepackage{amsmath}
\usepackage{algorithm,algorithmic}
 \restylefloat{algorithm}
\IEEEoverridecommandlockouts                              

\title{\LARGE \bf
Tracking Accuracy Based Generation Rules of Collective Perception Messages
}

\author{Shule LI$^{1}$ and Vincent Albert WOLFF$^{1}$
\thanks{*These authors contributed equally}
\thanks{$^{1}$Shule Li and Vincent Albert Wolff are with Institute of Communications Technology, Leibniz University Hannover.
       }%
}

\begin{document}

\maketitle
\thispagestyle{empty}
\pagestyle{empty}

\begin{abstract}

The Collective Perception Service (CPS) enables the enhancement of environmental awareness of Intelligent Transport System Stations (ITS-S) through the exchange of tracking information between stations. As the market penetration of CPS is growing, the intelligent distribution of the limited communication resources becomes more and more challenging. To address this problem, the ETSI CPS proposes dynamic-based object selection to generate Collective Perception Messages. However, this approach has limits and barely considers detection accuracy and the recent information available at other ITS-Ss in the transmission range. We show a proposal considering the current object tracking accuracy in the local environment and the object tracking from messages received by other stations in order to intelligently decide whether to include an object in a CPM. The algorithm decides based on the relative entropy between the local and V2X tracking accuracy if the object information is valuable for the nearby stations.  Our simulation according to the ITS-G5 standard shows that the Channel Busy Ratio (CBR) can be reduced with accuracy-based generation of CPM while improving Object Tracking Accuracy (OTA) compared to generation based on ETSI rules.

\end{abstract}

\section{INTRODUCTION}

In recent years, the number of vehicles equipped with Advanced Driver Assistance Systems (ADAS) has grown rapidly all over the world. One of the developments is to endow vehicles with sophisticated sensors and algorithms to perform advanced automated driving tasks. Vehicles use camera, LIDAR or radar sensors to observe the environment and fuse the information of all sensors to form their Local Environment Model (LEM). However, sensors may face some limitations, such as limited perception range, being occluded by other objects \cite{vehits19}, etc. In this case, enabling connectivity between vehicles by allowing them to exchange perception information helps overcome these limitations. 

Collective Perception (CP), designed to address the limitations mentioned above, is currently being standardized at ETSI (European Telecommunications Standards Institute) \cite{ITS2019}. According to the standard, Intelligent Transport System Stations (ITS-Ss) exchange the high-level information about the objects detected. CP transmits messages carrying a list of detected objects, including high-level information about their dynamic state (heading, speed, acceleration, etc.), dimension, confidence level, estimated type, and the detecting sensors. Various researchers focus on perception, suggesting that raw sensor data \cite{8885377} and deep features \cite{10.1145/3318216.3363300} should be exchanged. 
However, the transmission of raw sensor data and deep feature would require significant bandwidth, which can not be provided by current existing Vehicle to Everything (V2X) technologies such as ITS-G5, DSRC, C-V2X, etc. Therefore, this paper will rely on the exchange of CPM on the Facilities layer suggested by ETSI.

Collective Perception Message (CPM) format and CPM generation rules are defined in the Technical Report (TR) \cite{ITS2019}. These generation rules determine which object will be included in the next CPM and when the message will be sent. A study \cite{8813806} on TR shows that generation rules cause high redundancy and therefore could severely affect vehicular network congestion. This might be the case where multiple ITS-Ss detect the same object, and the generation rules will be triggered simultaneously on multiple stations causing unnecessary network channel load. Excess load would increase the risk of high transmission delays or even loss of valuable packets, potentially causing serious safety risks. This is because the current CPM generation rules are based exclusively on the result of local detection. Having a realistic sensor model, the detection may deviate from the ground truth because of the adverse weather condition (such as rain, snow, fog, etc.) or the occlusion by other obstacles, etc. The actual detection and object tracking may not follow the correct movement of the object but rather bounce around. This results in a frequent change of state and will generate more CPMs containing inaccurate detection, which "pollute" the LEM of nearby vehicles. Furthermore there is a negative impact on the channel load as unreliable data should not be sent out that frequently. ETSI added a confidence threshold to mitigate this effect.  

Another aspect is that the configuration of vehicle sensors differs. Some are more expensive and can provide high-precision detection results, while other vehicles are equipped with fewer or less expensive sensors and offer relatively low accuracy. CP encourages every ITS-S to share its detection results regardless of quality to expand the perception range. However, with more and more ITS-Ss participating in CP, some of them should reduce their dissemination frequency of CPMs to maintain an unimpeded communication channel. To ensure reliability of communication and road safety, each ITS-S should choose the transmitted data intelligently in order to save network resources and improve the knowledge of other ITS-S. 

We noticed that there is a lack of studies on generation rules of CPM considering local detection accuracy of ITS-Ss in their LEMs. In this paper, we propose an accuracy-based CPM generation rule that reduces the Channel Busy Ratio (CBR) and increases the Object Tracking Accuracy (OTA). We create an LEM for tracking the local detected objects and a V2X Environment model to track the objects which are received by CPM. Our method allows each ITS-S to intelligently select objects to be included in the next CPM by comparing the tracking accuracy of the LEM and the accuracy of the V2X Environment model. The object will be included if the local accuracy is greater than the accuracy of the V2X environment model with a predefined threshold. Our method adapts the idea proposed by \cite{8814110}, which introduces a cooperative perception framework that anticipates the value of an object for each vehicle. They showed a reduced tracking error through an analytical communication model. 
By extending the Artery framework, which couples OMNeT++ and SUMO \cite{krajzewicz2002sumo}, implementing the ETSI ITS-G5 \cite{etsi_itsg5} communication standard, we prove that our generation rules outperform the ETSI rules by reducing the CBR and the OTE.

\section{Related Work}
\subsection{ETSI Rules of CPM Generation}
According to the current ETSI CP message generation \cite{ITS2019}, the detection of the object will be included in the next CP message when it satisfies one of the following conditions: 
\begin{enumerate}
    \item The object is detected for the first time
    \item The absolute position of the object changes more than 4
    meters compared to the last time this object was included
    \item The absolute speed of the object changes more than 0.5 m/s compared to the last time this object was included
    \item The absolute heading of the object changes more than 4 degrees compared to the last time this object was included
    \item The object was not included in the last 1000 ms compared to the last time this object was included
\end{enumerate}

\subsection{Perception System}
Object detection remains a fundamental problem in the CP service and has been the focus of research in recent years in the area of deep learning. Various sensors like camera, LIDAR, etc. have been equipped on the vehicle to solve the above-mentioned problem. However, the camera sensor struggles with long distance measurement \cite{7353537} and provides a low accuracy result. An inherent drawback of LIDAR is the sparsity of point clouds in the long range and with partially occluded objects \cite{liang2018deep}. In this paper, we propose a sensor model in which the detection accuracy depends on the distance to the ego vehicle and the degree of occlusion.


\subsection{Object Tracking}
Detection generated by the on-board sensor, such as camera, radar, LIDAR should be aggregated using an association algorithm such as Joint Probabilistic Data Association Filter (JPDAF) \cite{5338565} and then tracked by an algorithm such as Kalman filter \cite{welch1995introduction}. Multi-object tracking is still a challenging topic, which is implemented in this paper by applying Kalman filters for the object tracking. We do not address the problem of data association and use a simplified object association approach.

\subsection{Information Filtering}
An ITS-S needs to be able to select and transmit the most valuable information to others to efficiently use the limited network resources. The author in \cite{aoki2020cooperative} presents a cooperative perception scheme with deep reinforcement learning to have connected vehicles to intelligently select the data. 

Using the concept of Value of Information (VoI) \cite{10.1145/2489253.2489265} to build the value anticipating network has also been used to prioritize messages in the vehicular network. The authors of \cite{5395777} propose to dynamically adjust the transmission power and time interval of beacon messages like Cooperative Awareness Message (CAM). In \cite{8814110}, the authors apply VoI to cooperative perception application and show an increased tracking performance as a result. This VoI is leveraged by ETSI CPS \cite{ITS2019} to define entropy-based redundancy mitigation rules.

\section{Motivation}
The generation rules of CPM studied by ETSI and in many other papers \cite{8813806} \cite{9062827} take object detection for granted and focus on communication. However, overlooking object tracking accuracy may cause a potential risk when the ETSI dynamic rules are implemented. These rules should be triggered by object motion. However, if the tracking state of the object changes frequently, the object will be included in the CPM more often, as the ego vehicle perceives the object as moving. The change in detection could be correctly caused by the object changing its state very fast, but  also because of low accuracy of the on-board sensors. The latter could lead to significant issues in detecting objects with low accuracy and result in frequent transmission of inaccurate object states. In this paper, the focus is on CPM generation rules considering detection and tracking accuracy.

The vehicles we study are Connected and Automated Vehicles (CAVs) which are equipped with on-board sensors and a communication unit. In reality, the result generated by different vehicles varies greatly depending on their sensing system. In addition, the same vehicle will generate perception with different accuracy when detecting an object under different conditions. It may seem obvious to decide that detection with high accuracy should be prioritized and sent first. The lack of an omniscient view makes it challenging to determine whether the detection of the ego vehicle is better than detection by others. We will discuss three cases at an intersection and try to understand a reasonable way of transmitting CPMs.
   \begin{figure}[thpb]
      \centering
        \includegraphics[scale=0.56]{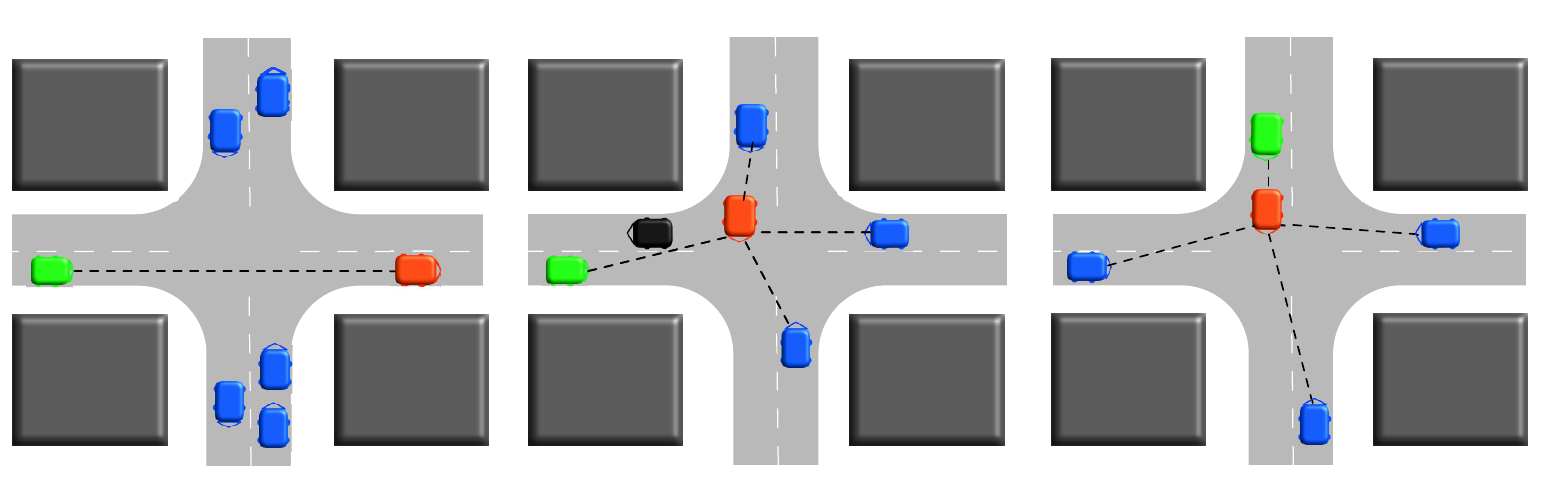}
        (a) \quad\quad \quad \quad \quad \quad \quad  (b) \quad\quad \quad \quad \quad \quad \quad  (c)

      \caption{Example scenarios whether an object should be sent in the CPM }
      \label{3cases}
   \end{figure}

In figure \ref{3cases}, the black and red vehicles do not have a communication unit, while the other vehicles are all CAVs. We consider the green vehicle as the ego vehicle and the red vehicle as the object to be detected. The dot lines refer to a successful detection. In case (a), only the ego vehicle is capable of detecting the object, no matter the detection accuracy, the object should be included in the next CPM of the ego vehicle. In case (b), the ego vehicle detects the object with low-accuracy detection due to the black vehicle. In the meantime, other CAVs also detect the object. In this case, the ego vehicle may choose not to include the object in the next CPM as the others may have a more accurate detection. In case (c), the ego vehicle detects the object with a higher accuracy than the other vehicles considering that the distance to the object is the closest. So in this case, the ego vehicle should include the object detection in the next CPM.

The ego-vehicle needs to be aware of how accurately it measures the object and how the others measure the same object in order to decide whether to include its detection. When the ego-vehicle performs the detection process of an object, a confidence level is given along with the measurement, which shows how well the ego-vehicle estimates this measurement. Currently in the ETSI CPS, the confidence level is defined as a scalar value from 0 to 1. A scalar value may fit a single measurement, but normally a series of measurements is going to be performed, and the detection should be accumulated in a specific way called tracking. In this case, the scalar number is not sufficient to represent the tracking, and we need to find another metric to quantify its goodness. The Kalman filter is widely used as a tracking algorithm, which maintains an updated covariance matrix after each measurement. This represents the tracking confidence level more precisely than a scalar value. The ego vehicle also tracks the detection from other vehicles through V2X communication about the same object. Then, by comparing these two tracking, the ego vehicle can make a decision about whether to include this measurement in the next CPM. Relative information entropy can be used to compare the local knowledge such as the measured absolute state and the knowledge gained by V2X communication.

\section{Object Tracking Accuracy Based CPM Generation}

\subsection{Sensor Model}
\begin{figure}[thpb]
      \centering
        \includegraphics[scale=0.175]{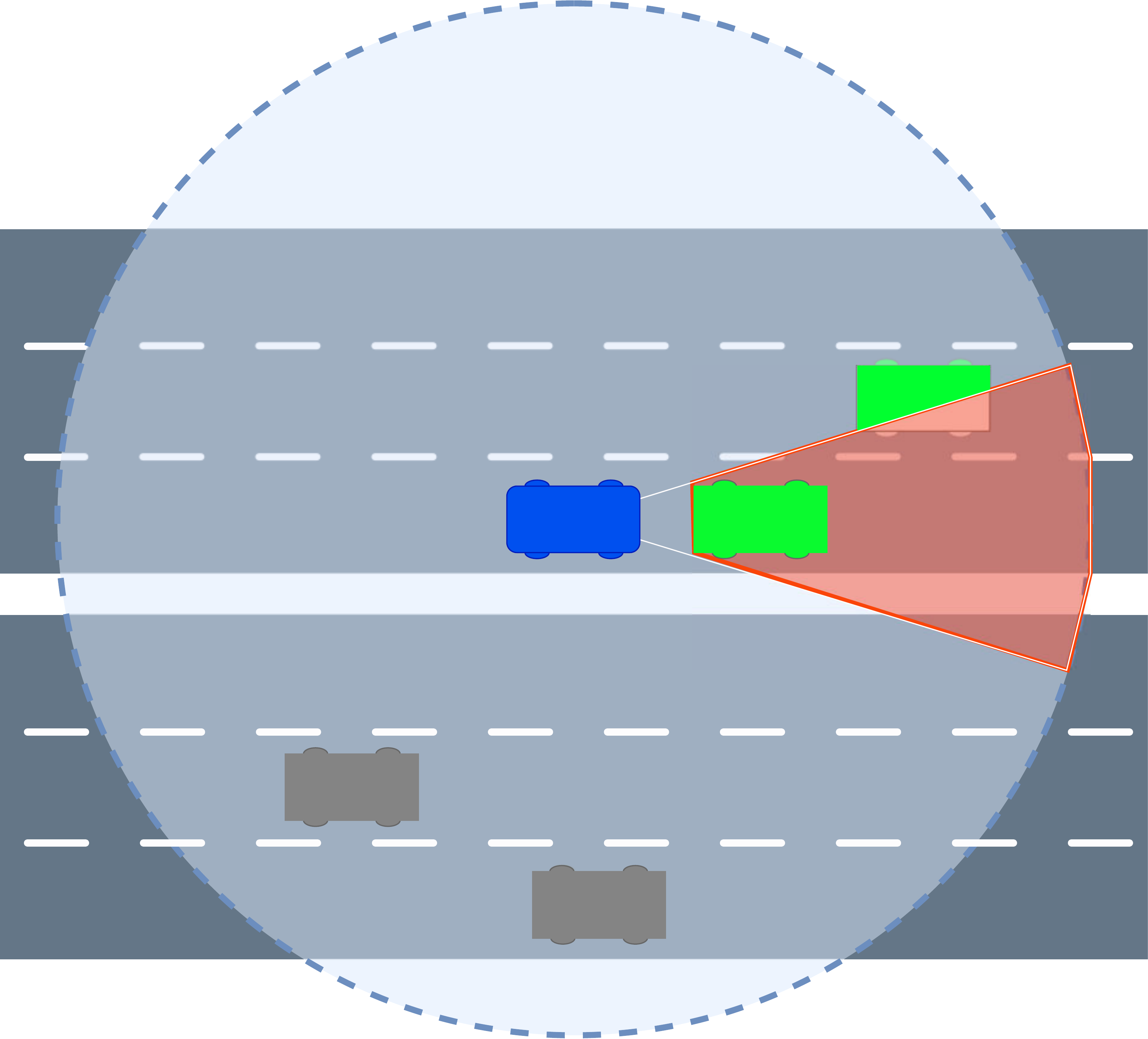}
      \caption{Sensor model in a highway scenario}
      \label{sensor_model}
   \end{figure}

In order to obtain information about surrounding objects, vehicles are equipped with a 360-degree camera that has a detection range of up to 85 m, with a frequency of 10Hz. Figure \ref{sensor_model} illustrates the sensor model with an example of an occlusion: The ego vehicle (blue) is partly detecting the object on the second lane, which shrinks the visible cross-section area (green) and causes an occluded area (red). The vehicle in front is fully detected. \\Sensor detection is realized with a probabilistic model: Based on the visible cross-section, defined as the surface of the detected object that is not occluded an in sensor range, and the relative distance to the ego vehicle of a detected object, the measurement accuracy is determined by adding noise to an object's ground truth position. The detection accuracy $\mathcal{D}$ of an object $\mathcal{O}$ is therefore a function $\mathcal{D}_{\mathcal{O}}(c,d) = (\sigma_{x},\sigma_{y})$, with parameters cross-section $c$ and distance $d$. The function returns a tuple $(\sigma_{x},\sigma_{y})$, which corresponds to the variance of the Cartesian coordinates of the detection. After the variance is determined, the ground truth value of the $(x,y)$ position is taken as the mean value $\mu$ of the normal distribution $\mathcal{N}_{(\sigma,\mu)}$. A random value from this distribution is selected as the current sensor measurement $\mathbf{z}_{k}$.
\subsection{Kalman Filter}
The described Gaussian distributed detection model reveals a high scatter of the detection positions, due to the characteristic of memorylessness of a Gaussian random process. Therefore, a Kalman filter is applied that attenuates the impact of the current measurement on the tracking position by continuously applying a dynamic vehicle model adjusting the sensor measurement against the previous vehicle track. The current detection $\mathcal{D}$ updates the Kalman filter for every incoming detection update at time step $k$. This update process is divided into the prediction and the update step.
$$
\mathbf{x}_{k} = \mathbf{x}_{k-1} \cdot \mathbf{F}_{k-1}\eqno{(1)}
$$
As depicted in equation (1) the state vector $\mathbf{x}_{k}$ is predicted based on the previous state $\mathbf{x}_{k-1}$ and the matrix $\mathbf{F}$, which describes the dynamic motion model of the vehicles. In our case, we use a constant-velocity motion model.
$$
\mathbf{P}_{k} = \mathbf{F}_{k-1} \cdot \mathbf{P}_{k-1} \mathbf{F}^{T}_{k-1} + \mathbf{Q}_{k-1}
\eqno{(2)}
$$
As in the next step (2), the error covariance matrix $\mathbf{P}$ is updated. Here, the process noise $\mathbf{Q}$ is added. 
After the prediction is done, the update step is initiated, which corrects the prediction based on the sensor measurement by calculating the Kalman gain $\mathbf{K}$. Equation (3) shows that sensor measurement noise is taken into account, which is represented as the covariance matrix $\mathbf{R}$. As the measurement itself, the noise depends on cross-section and distance. In (4) the state update $\mathbf{x}_{k}$ is performed based on the current sensor measurement $\mathbf{z}_{k}$. The process covariance matrix is updated as stated in (5). \cite{5512258}
$$
\mathbf{K}_{k} = \mathbf{P}_{k} \mathbf{H}^{T}_{k} \cdot (\mathbf{H}_{k} \mathbf{P}_{k} \mathbf{H}^{T}_{k} + \mathbf{R}_{k})^{-1}
\eqno{(3)}
$$
$$
\mathbf{x}_{k} = \mathbf{x}_{k-1} + \mathbf{K}_{k} (\mathbf{z}_{k} - \mathbf{H}_{k} \mathbf{x}_{k})
\eqno{(4)}
$$
$$
\mathbf{P}_{k} = (\mathbf{I} - \mathbf{K}_{k} \mathbf{H}_{k}) \mathbf{P}_{k}
\eqno{(5)}
$$
\subsection{Kullback-Leibler divergence}
\label{kullback-leibler}
To estimate the similarity of probabilistic distributions, in our case different covariance matrices $\mathbf{P}$, a proper metric has to be applied. The Kullback-Leibler divergence \cite{kullback1951information} measures the relative entropy of two probabilistic distributions. It should not be confused with a distance measurement, as the Kullback-Leibler divergence is non-symmetric. In our case, the probability distributions of local tracking $\mathcal{N}_{LEM}$ represented by $\mathbf{P}_{LEM}$ and V2X tracking with enabled communication $\mathcal{N}_{LEM}$ represented by $\mathbf{P}_{V2X}$ are compared according to the equation stated in (6). 

\begin{equation}
\begin{split}
\setcounter{equation}{6}
D_{KL}(\mathcal{N}_{LEM} \parallel \mathcal{N}_{V2X})  =
\frac{1}{2}(\log{ (\frac{\det{\mathbf{P}_{V2X}}}{\det{ \mathbf{P}_{LEM}}})} - k \\ +   tr(\mathbf{P}_{V2X}^{-1} \mathbf{P}_{LEM}) + (\mu_1 - \mu_0)^\mathsf{T} \mathbf{P}_{V2X}^{-1} (\mu_1 - \mu_0))
\end{split}
\end{equation}
\subsection{Environment Model}
For each vehicle, three different environment models should be tracked. Local Environment Model (LEM) is built based on the detection of the on-board sensor with the Kalman filter $\mathcal{K}_{LEM}$ for each perceived object. To track the objects included in the received CPM messages, a V2X Environment Model with the Kalman filter $\mathcal{K}_{V2X}$ for each object in received CPM is introduced. These two Environment Models track objects independently, although the joint intersection of tracked objects is supposed to be big. As ITS-Ss fuse received object information and local tracked objects to minimize tracking error, a third Kalman filter $\mathcal{K}_{Fused}$ is created for each object, which combines information from the LEM and the V2X model. 

 \begin{algorithm}
 \begin{algorithmic}[1]
 \renewcommand{\algorithmicrequire}{\textbf{Input:}}
 \renewcommand{\algorithmicensure}{\textbf{Output:}}
 \REQUIRE $\mathbf{K}_{LEM}$, $\mathbf{K}_{V2X}$
 \ENSURE CPMObjectList
  \FOR {Tracking $\mathbf{x}_{k}$ in LEM}
  \IF {(trace($\mathbf{P}_{LEM}$) $<$ $\mathcal\theta$)}
    \IF{($D_{KL}$($\mathbf{P}_{LEM}$ $\parallel$ $\mathbf{P}_{V2X}$) $>$ $\mathcal \gamma$)}
        \STATE CPMObjectList add $\mathbf{x}_{k}$ 
    \ENDIF
  \ENDIF
  \ENDFOR
  \RETURN CPMObjectList
 \end{algorithmic} 
  \caption{Algorithm for selecting objects in the next CPM}
 \end{algorithm}
\subsection{Proposal}
The goal of our proposal is to intelligently select which object should be included in the next CPM. To this aim, the algorithm should be triggered every 100 ms, which corresponds to the highest CPM frequency suggested by ETSI. The inputs $\mathbf{K}_{LEM}$ and $\mathbf{K}_{V2X}$ are two lists of the Kalman filter $\mathcal{K}_{LEM}$ and $\mathcal{K}_{V2X}$. For each LEM tracking, the covariance matrix $\mathbf{P}_{LEM}$ should be checked to make sure that the local tracking works correctly. Comparing the trace of $\mathbf{P}_{LEM}$ and a pre-defined threshold $\mathcal\theta$, if the trace is smaller than the threshold, then the tracking is considered valuable. In our case, we define $\mathcal\theta$ = 1. Having implemented these separate filters, the ITS-S can compare its local tracking with the V2X tracking based on the Kullback-Leibler divergence stated in Chapter \ref{kullback-leibler} by calculating the relative entropy of $\mathbf{P}_{LEM}$ and $\mathbf{P}_{V2X}$. Based on the result, a threshold $\mathcal \gamma$ is defined for comparison with the relative entropy. If the relative entropy is sufficiently high, this tracking should be included in the next CPM as it contributes significantly to the LEM of surrounding ITS-Ss. The decision how to set the value for $\mathcal \gamma$ will be discussed after testing different values in the simulation. The pseudo-code for this process is shown in Algorithm 1.

\begin{table}[t]
    \centering
    \begin{tabular}{l|l}
         \textbf{Parameter} & \textbf{Value}  \\
         Physical layer & IEEE 802.11p\\
         \hline
         Bit rate & 6Mbit/s \\
         \hline 
         Carrier frequency & 5.9GHz \\
         \hline
         Bandwidth & 10MHz \\
         \hline
         Channel & G5-CCH \\
         \hline
         Transmission power & 200 mW \\
         \hline
         Signal threshold & -85 dBm \\
         \hline
         Noise threshold & -65 dBm \\
         \hline
         V2X propagation model & GEMV2 \cite{Boban2014GeometryBasedVC}
         \\
         \hline
         Vehicle dentity &
         \begin{tabular}[c]{@{}l@{}}60Veh/km (low)\\120Veh/km (high)\end{tabular}  \\
         \hline
         Penetration rate & 100\% \\
         \hline
         Decentralized Congestion Control & Disabled\\
    \end{tabular}
    \caption{Simulation parameters}
    \label{tab:sim}
\end{table}
\section{Simulation}
For simulation the Artery framework \cite{artery2015} is used, providing an implementation of ETSI ITS-G5 \cite{etsi_itsg5} communication. The simulation is built with parameters according to table \ref{tab:sim}. Artery is coupled with SUMO \cite{krajzewicz2002sumo}, which provides a microscopic traffic simulation. In SUMO we create a straight highway scenario with 3 lanes in both directions according to figure \ref{sensor_model}, with a total length of 5 km. The idea of this scenario was taken from \cite{8813806} to generate results under low and high congested communication channel. 
Vehicles will enable communication only when entering an area of 2 km in length to restrict the number of nodes generated, which speeds up the simulation process. The data logging area is limited as well, it is enabled only in the middle 1 km part to avoid boundary effects. The communication of all simulated vehicles within this area is evaluated. The market penetration rate of CPM is set to 100\%, so all stations are capable of CPM transmission. To analyze the impact of network congestion, two different vehicle densities are simulated: one low-density scenario (60 Veh/km) and one high-density scenario (120 Veh/km).

\section{Evaluation}
\begin{figure}[t]
      \centering
        \includegraphics[scale=0.52]{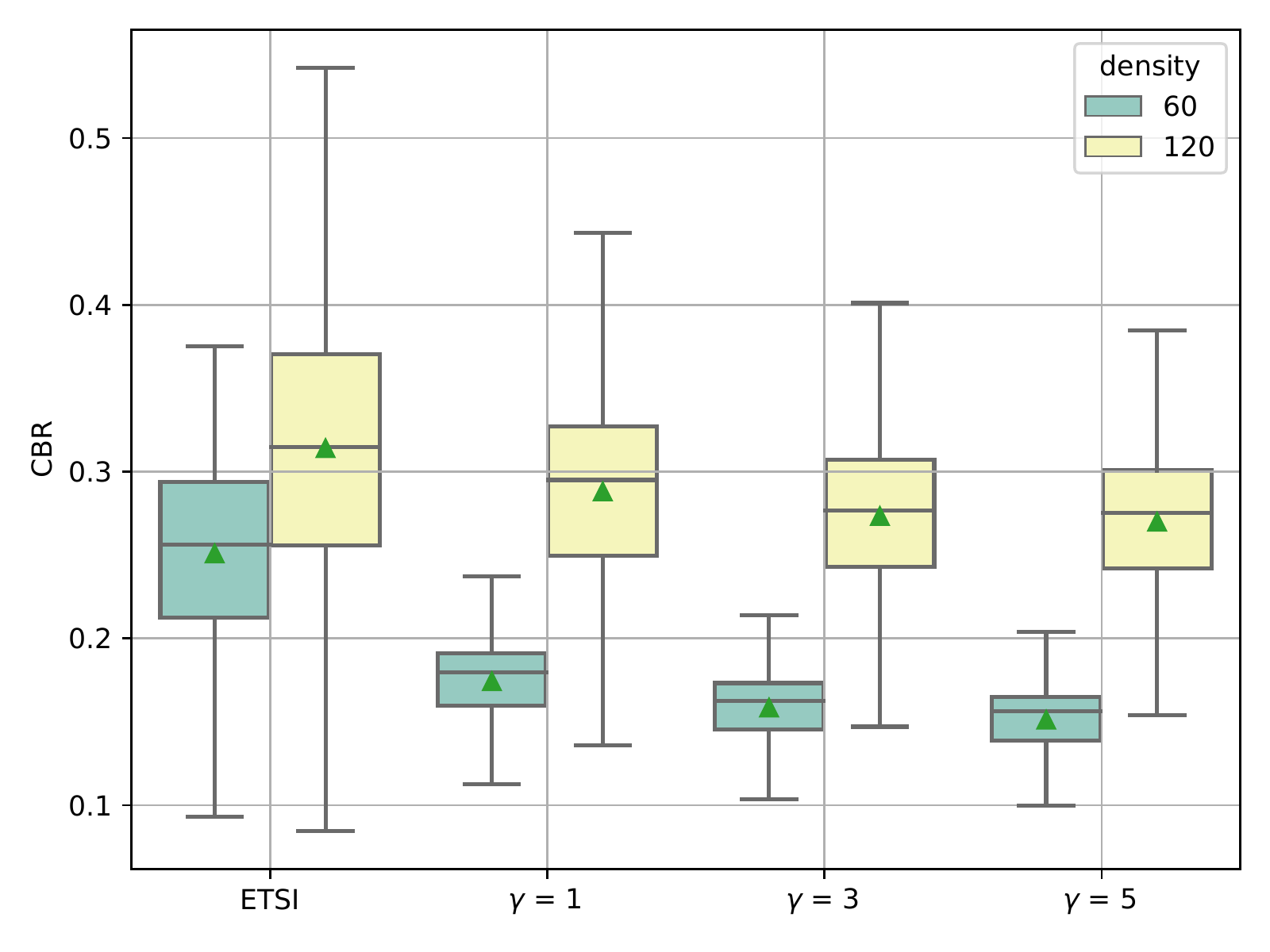}
      \caption{CBR for low traffic scenario and high traffic scenario}
      \label{cbr}
   \end{figure}

In this section we present the results generated with the simulation setup mentioned in the previous chapter for the ETSI CPM generation rules and our method. Using our method, the result are generated with different thresholds $\gamma$. 
Figure \ref{cbr} shows the Channel Busy Ratio (CBR) for low traffic density (60 vehicle per kilometer) and high traffic density situation (120 vehicle per kilometer). For low density scenario, with threshold set to 1 the CBR is reduced by 32\% (from 25\% to 17\%). When the threshold is increased, the CBR still decreases but without a big margin. The mean CBR for $\gamma$ = 3 and $\gamma$ = 5 are both around 15\%, which is a relative low number where packet collisions are rarely to be expected. For high density scenario, the CBR decreases following the same pattern as in the low density case. Although there is no significant difference in the mean value, we can still see that the high CBR cases are mostly omitted and values larger than 0.5 are rarely reached. The channel congestion with our filtering method is expected to be lower than the ETSI-based approach because more low-accuracy detection is omitted.
In our simulation, we only enable the vehicles to generate CPM, but in reality, vehicle will also send other messages like Cooperative Awareness Message (CAM) for beaconing or Decentralized Environmental Notification Message (DENM) for the alert of safety related events. Therefore, the channel will become more congested and the reduction of CBR will be beneficial to disseminate different types of V2X messages.

\begin{figure}[t]
      \centering
        \includegraphics[scale=0.52]{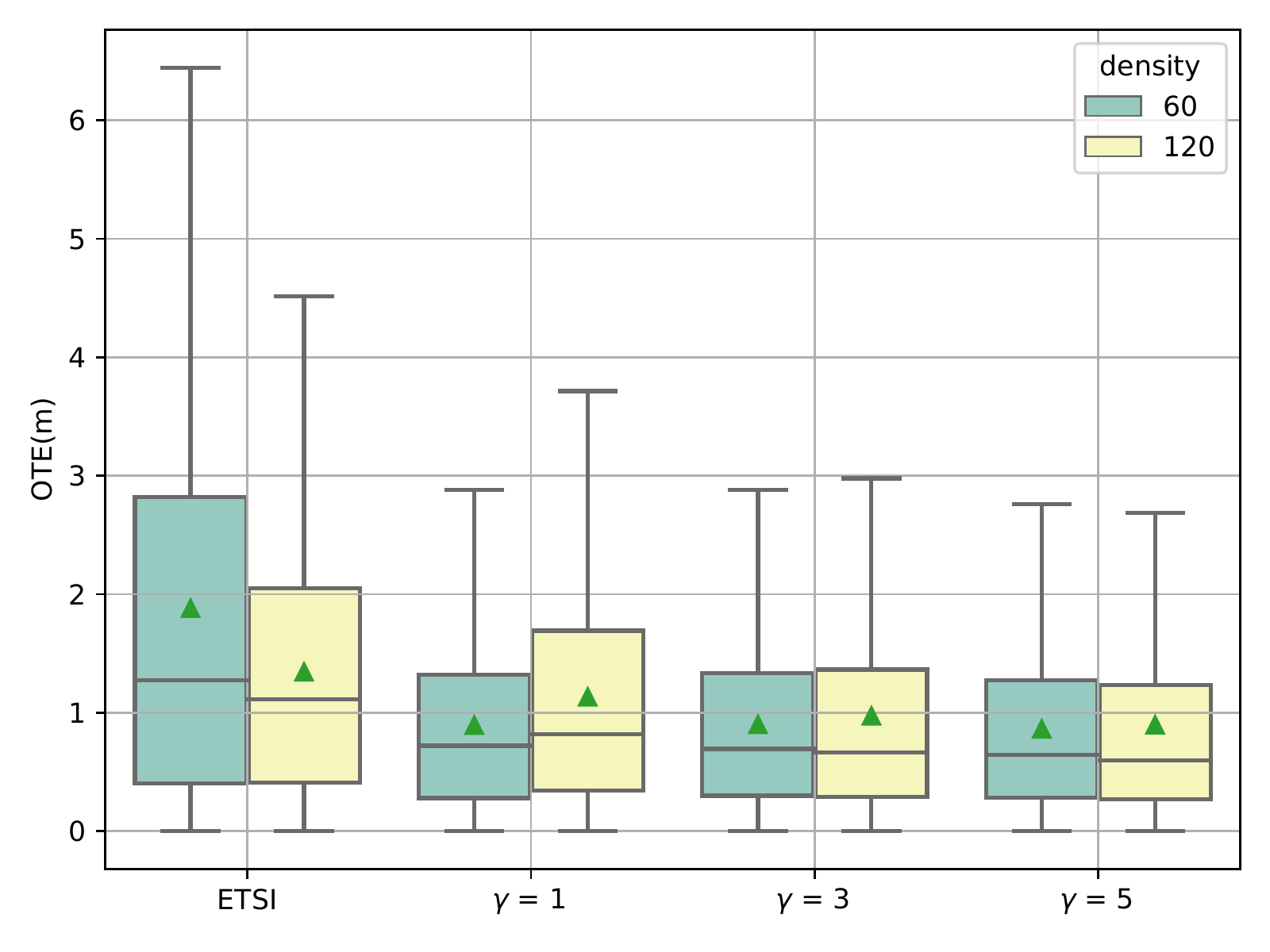}
      \caption{OTE for low traffic scenario and high traffic scenario}
      \label{ota}
   \end{figure}

Another important aspect with which we want to analyze our method is the evaluation of perception. Object tracking error (OTE) is defined as the distance between the object ground truth position and the tracking position from the ego vehicle. Previous studies \cite{8813806} \cite{9062827} focus only on the Environmental Awareness Ratio (EAR), since perfect tracking is assumed for their perception model, which makes logging OTE not feasible in the simulation. 
In figure \ref{ota} we see that in the low traffic density scenario, the upper whisker tracking error is more than 6 meters and the mean, indicated by a triangle in the box plot, is around 2 meters, which impacts the reliability of object tracking significantly. With our method and $\gamma$ set to 1, the upper whisker of OTE is reduced to 3 meters and the mean tracking is reduced by about 55\%. When we increase the threshold, the tracking does not change significantly. On the other hand with high density, if we increase $\gamma$ from 1 to 3, we can notice the tracking accuracy is improving and the maximum error is reduced. The results show that our method can improve the tracking accuracy of the objects by a big margin compared to ETSI. 

If we consider both the CBR and OTE, implementing our method improves the CBR while increasing the accuracy of tracking. With our sensor model, when ego vehicle detects an object from far away or the object is partly occluded, the detection result contains a relatively big error. This error will become worse if we are using ETSI rules: In this case, the detection is more likely to trigger the generation of CPMs. For this reason, tracking errors of up to above 6m are shown in figure \ref{ota}. The result proves that the detections which will generate large tracking errors are dropped according to our method. Also, we notice that with increasing $\gamma$ from 3 to 5, the CBR as well as OTA does not change significantly. It seems that in our highway scenario, an increase in threshold from 3 to 5 has less benefit because bad tracking is already filtered out. So we will keep $\gamma$ = 3 and investigate the relation between tracking error and distance.

\begin{figure}[t]
      \centering
        \includegraphics[scale=0.52]{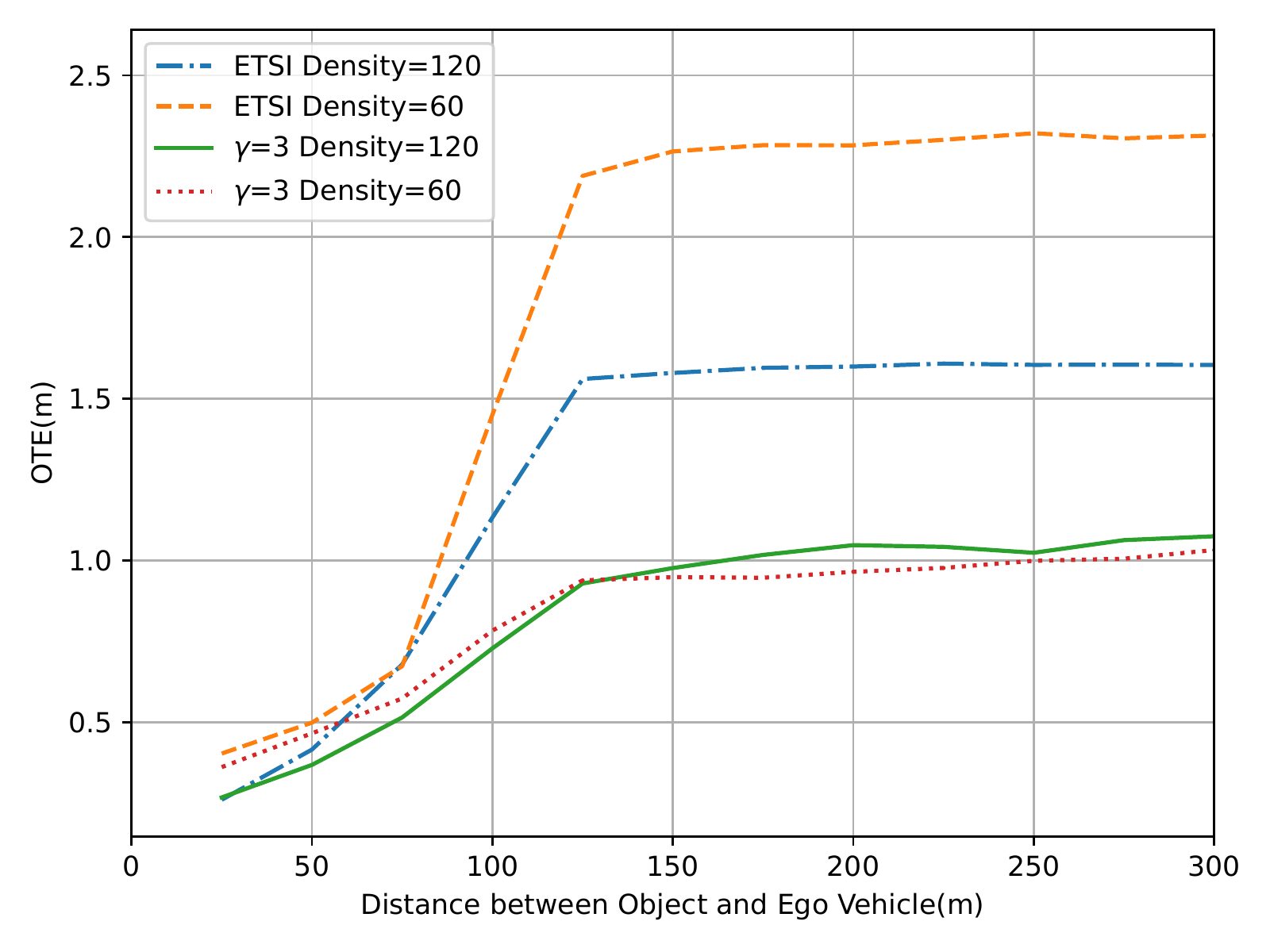}
      \caption{Mean OTE for $\gamma$ = 3 compared with ETSI}
      \label{ota_distance}
   \end{figure}

In figure \ref{ota_distance}, the mean tracking error based on the distance from ego vehicle to object is plotted. The tracking errors for all methods are increasing rapidly until a distance to the object of 125 meters and slowly increasing until 300 meters. We only show the value until 300 meters because this range is considered as a critical distance. The on-board sensor model has a detection range of 85m, which provides accurate tracking within this range. From 85m on, the vehicles have to rely on CPM sent by the nearby stations and we notice a big increase of the tracking error for all simulation parameters. However with our method, the bad detection is highly attenuated so that the tracking error decreases from 1.5m for high-density and 2.3m for low-density to lower than 1 meter for each. This again confirms that the object selection of our method is superior, which is beneficial for the whole vehicular network. It is also interesting to notice that the tracking error is lower if we compare high density to low density. This is mainly caused by two differences: In the high density scenario, vehicles are moving slower due to the higher road congestion and their inter-gap is smaller between each other, so the tracking error is smaller. Another aspect is that information about one object is received more frequently, as more vehicles are tracking the same object. As seen in figure \ref{ota}, with our method the average tracking error is less than 1 meter for all densities and set thresholds. This shows that our method improves the object tracking significantly and leads to an improvement of vehicular safety.

\section{Conclusions and Future Work}
Taken into consideration a sensor model that is based on detection accuracy and introducing a CPM generation rule based on the object tracking accuracy, which compares the LEM with the V2X object tracking, our method shows significant improvement in terms of tracking error (OTE) and channel load (CBR) compared to the ETSI CPM generation rules. Decreasing the tracking error while reducing the CBR improves the reliability of information received from the V2X network and can lead to an improvement of vehicle safety. More network resources can be granted to other communication services such as CAM. \\

In future work, the simultaneous run of different V2X services such as CAM, DENM and CPM could be considered to prove the benefits of intelligent CPM rules to the overall network performance. Also the coupling of our method and the DCC on Access or Facilities layer may show the impact of our object filtering on different congestion control implementations. 

\addtolength{\textheight}{-12cm}   


\section{Acknowledgements}
This publication is partially funded by the Lower Saxony Ministry of Science and Culture under grant number ZN3493 within the Lower Saxony “Vorab“ of the Volkswagen Foundation and supported by the Center for Digital Innovations (ZDIN). This publication is also partially funded by Deutsche Forschungsgemein-schaft (DFG, German Research Foundation) - project number 227198829 / GRK1931.
\bibliographystyle{ieeetr}
\bibliography{References.bib}



\end{document}